\documentstyle[seceq,epsf,twocolumn]{jpsj}
\author{Yoshiki {\sc Imai} and Norio {\sc Kawakami}}
\title{Heavy Quasi-Particles in the Two-Orbital Hubbard Model}
\inst{Department of Applied Physics, 
Osaka University, Suita, Osaka 565-0871}

\recdate{}

\abst{The two-orbital Hubbard model with the Hund 
coupling is investigated in a 
metallic phase close to the Mott insulator.
We calculate the one-particle spectral function and the optical conductivity 
within dynamical mean field theory, 
for which the effective impurity problem 
is solved by using the non-crossing approximation. 
For a metallic system close to quarter filling, a heavy quasi-particle 
band is formed by the Hubbard interaction, the effective mass of which
is not so sensitive to the orbital splitting and the Hund coupling.
In contrast, a heavy quasi-particle band near 
half filling disappears in the presence of the orbital splitting, but
is induced again by 
the introduction of the Hund coupling,  resulting in
a different type of heavy quasi-particles.
}

\kword{strongly correlated electron systems, Hund coupling, orbital splitting,
dynamical mean field theory}

\def\runtitle{Heavy Quasi-Particles in the Two-Orbital Hubbard Model} 
\def\runauthor{Yoshiki {\sc Imai} and Norio {\sc Kawakami}}

\begin{document}
\sloppy
\maketitle

%%%%%%%%%%%%%%%%%%%%%%%%%%%%%%%%%%%%%%%%
\section{Introduction}
%%%%%%%%%%%%%%%%%%%%%%%%%%%%%%%%%%%%%%%%
Strongly correlated electron systems, 
such as  the high-$T_{\rm c}$ superconductors \cite{Ima98} 
and the heavy fermion compounds \cite{Hew93}, 
have attracted much interest.
In order to deal with electron correlations,
the single-orbital Hubbard model \cite{Gut63,Kan63,Hub64}
has been studied extensively with various analytical and numerical methods.
%%%%%%%%%%%%%%%%%%%%%%%%%%%%%%%%%%%%%%%
Among others, dynamical mean field theory (DMFT) \cite{Geo96}
enables us  to systematically investigate  strong 
correlation effects.
By using  DMFT
characteristic properties related to the Mott transition 
 has been described successfully for the single orbital case.\cite{Geo96}

In real materials such as the transition metal oxides,\cite{Ima98}
orbital degeneracy plays an 
important role, giving rise to a wide variety 
of interesting phenomena. In such cases the Hund coupling 
and  the orbital splitting are the additional key parameters
 to discuss electron correlations.
It is thus desirable to study the orbitally degenerate  Hubbard model
in order to better understand
 strongly correlated electron systems.
Recently, a number of theoretical studies on the 
orbitally degenerate Hubbard model have been done by using 
the slave-boson mean field approximation \cite{Has97,Fre97,Kle98},
the quantum Monte Carlo  method \cite{Mot98} 
and DMFT \cite{Kot96,Roz97,Han98,Mom98,Lom00}.

In this paper, we systematically study electron correlations 
for the two-orbital Hubbard model with particular 
emphasis on the interplay of 
the Hubbard interaction, the Hund coupling and the orbital splitting. 
By employing DMFT combined with the non-crossing approximation (NCA),
we compute the one-particle spectral function and the optical conductivity. 
We discuss how the heavy quasi-particle behavior shows up
in a metallic state close to the Mott insulator 
with orbital degeneracy.  
We find that the origin of heavy quasi-particles can be 
different between quarter filling and half filling. 
 In particular, we point out that 
 the Hund coupling plays an essential role to  
form a heavy quasi-particle band in the vicinity of half filling.

This paper is organized as follows. In the next section, we 
briefly describe the model and the method, and then in \S 3 we 
show the results obtained for the one-particle spectral function
and the optical conductivity.  
A brief summary is given in \S 4.

%%%%%%%%%%%%%%%%%%%%%%%%%%%%%%%%%%%%%%%%
\section{Model and Method}
%%%%%%%%%%%%%%%%%%%%%%%%%%%%%%%%%%%%%%%%

%=======================================
\subsection{Two-Orbital Hubbard Hamiltonian}
%=======================================

We study the Hubbard model with two orbitals. 
The Hamiltonian is given by
%%%%%%%%%%%%%%%%%%
\begin{eqnarray}
H&=&\sum_{<i,j>}\sum_{m,m',\sigma}t^{mm'}_{ij}c^{\dag}_{im\sigma}c_{jm'\sigma}
+\frac{\Delta}{2}\sum_{i,\sigma}(n_{i1\sigma}-n_{i2\sigma})\nonumber \\
&+&U'\sum_{i,m}n_{im\uparrow}n_{im\downarrow}
+U\sum_{i,\sigma}n_{i1\sigma}n_{i2\bar{\sigma}}\nonumber \\
&+&(U-J)\sum_{i,\sigma}n_{i1\sigma}n_{i2\sigma}\nonumber \\
&-&J\sum_{i}(c^{\dag}_{i1\uparrow}c_{i1\downarrow}
c^{\dag}_{i2\downarrow}c_{i2\uparrow}+h.c.),
\end{eqnarray}
%%%%%%%%%%%%%%%%%%%%%%%%%%%%
where $c_{im\sigma} (c^{\dag}_{im\sigma})$ is the annihilation 
(creation) operator of electrons with the orbital $m(=1,2)$ and 
spin $\sigma(=\uparrow,\downarrow)$ at the site $i$ and 
$\Delta$ represents the energy  splitting between the two orbitals. 
Here, $U' (U)$ is the Coulomb repulsion 
between electrons in the same (different)
orbital and $J$ is the ferromagnetic Hund coupling. 
When the two orbital-levels are degenerate ($\Delta=0$), 
the system is rotationally invariant with respect to the
spin and orbital degrees of freedom, leading to
 the condition, $U'=U+J$. 
We use this relation and assume that the hopping
 is diagonal with respect to the orbital indices
($t^{mm'}_{ij}=-t\delta_{mm'}$).  We shall deal with
a paramagnetic metallic phase close to the Mott insulator
in this paper.

%=======================================
\subsection{Dynamical Mean Field Theory}
%======================================

It is known that 
DMFT is justified in the limit of 
large spatial dimensions
\cite{Met89,Mul89} 
and gives a rather good approximation even in three dimensions. 
In this method, a lattice electron problem is replaced by 
the corresponding impurity one 
embedded in an effective medium determined self-consistently.
%%%%%%%%%%%%%%%%%%%%%%%%%%%%%%%%%%%%%%%%%%%
The effective local  Hamiltonian is given by 
%%%%%%%%%%%%%%%%%%%%%%%%%%%%%%%%%%%%
\begin{eqnarray}
H_{\rm eff}&=&H_{\rm loc}+H_{\rm med},\\
H_{\rm loc}&=&\sum_{m,\sigma}E_{fm}n_{m\sigma}
+U'\sum_{m}n_{m\uparrow}n_{m\downarrow}\nonumber \\
&+&U\sum_{\sigma}n_{1\sigma}n_{2\bar{\sigma}}
+(U-J)\sum_{\sigma}n_{1\sigma}n_{2\sigma}\nonumber \\
&-&J(c^{\dag}_{1\uparrow}c_{1\downarrow}
c^{\dag}_{2\downarrow}c_{2\uparrow}+h.c.), \\
H_{\rm med}&=&\sum_{km\sigma}(V_{km\sigma}a^{\dag}_{km\sigma}c_{m\sigma}+h.c.)
\nonumber \\
&+&\sum_{km\sigma}\epsilon_{k}a^{\dag}_{km\sigma}a_{km\sigma},
\end{eqnarray}
%%%%%%%%%%%%%%%%%%%%%%%%%%%%%%%%%%%%%%%%%%%%%
where the impurity level in the reduced problem is 
%%%%%%%%%%%%%%%%%%%%%%%%%%%%%%
\begin{eqnarray}
E_{{f}m}=\left \{
\begin{array}{cc}
E_{f}+\Delta/2 & ({\rm orbital\,1}) \\
E_{f}-\Delta/2 & ({\rm orbital\,2})
\end{array}
\right.
\end{eqnarray}
%%%%%%%%%%%%%%%%%%%%%%%%%%%%%%%%%%%%%%%%%%%%%%%%%%%%%
Here, $V_{km\sigma}$ represents  the hybridization between the impurity 
site and an effective medium. 

Let us first introduce the
 so-called {\it cavity} Green function and the {\it full} Green function, 
which are defined respectively as
%%%%%%%%%%%%%%%%%%%%%%%%%%%%%%%%%%%%%%%%
\begin{eqnarray}
{\cal G}_{0m\sigma}(\omega)&=&
[\omega +{\rm i}\delta +\mu -E_{fm} -\Delta_{m\sigma}(\omega)]^{-1}\\
G_{m\sigma}(\omega)&=&
[{\cal G}_{0m\sigma}^{-1}(\omega)-\Sigma_{m\sigma}(\omega)]^{-1}
\end{eqnarray}
%%%%%%%%%%%%%%%%%%%%%%%%%%%%%%%%%%%%%%%%%%%%%%
where $\mu$ is the chemical potential. 
Here  $\Delta_{m\sigma}(\omega)$ represents 
the hybridization function given by
%%%%%%%%%%%%%%%%%%%%%%%%%%%%%%%%%%%%
\begin{eqnarray}
\Delta_{m\sigma}(\omega)=\frac{1}{N}\sum_{k}
\frac{|V_{km\sigma}|^{2}}{\omega +{\rm i}\delta -\epsilon_{k}}.
\end{eqnarray}
%%%%%%%%%%%%%%%%%%%%%%%%%%%%%%%%%%%%%%%%%%%%%%%%
In  DMFT, the self-consistency equation is given by
%%%%%%%%%%%%%%%%%%%%%%%%%%%%%%%%%%%%%%%%%%%%%%%%%%%%%%%
\begin{eqnarray}
&&G_{m\sigma}(\omega)= \nonumber \\
&&[\omega +{\rm i}\delta +\mu -E_{fm}-\Sigma_{m\sigma}(\omega) 
-\frac{D^{2}}{4}G_{m\sigma}(\omega)]^{-1},
\end{eqnarray}
%%%%%%%%%%%%%%%%%%%%%%%%%%%%%%%%%%%%%%%%%%
where we have employed the semi-elliptic density of states 
 for non-interacting electrons,
$N_{0}(\epsilon)=2/(\pi D^{2})\sqrt{D^{2}-\epsilon^{2}}$ 
with the bandwidth $D$.

In order to solve the effective impurity problem, 
we further use NCA\cite{Pru93,Kei70,Kur83,Col84,Mul84,Bic87,Pru89}, 
which is known to provide rather reliable results in agreement with 
the quantum Monte Carlo
calculation \cite{Pru93} for the single-orbital case. 
We first diagonalize the local Hamiltonian $H_{\rm loc}$, 
%%%%%%%%%%%%%%%%%%%%%%%%%%%%%%%%%%%%%%%%%%%%%%
\begin{eqnarray}
H_{\rm loc}=\sum^{16}_{\alpha=1}E_{\alpha}|\alpha \rangle \langle \alpha|,
\end{eqnarray}
%%%%%%%%%%%%%%%%%%%%%%%%%%%%%%%%%%%%%%%%%%%%%%%%%%%%
where $|\alpha \rangle$ and $E_{\alpha}$ 
denote the spin-orbital eigenstate and the corresponding 
energy for the isolated ionic system. 
Using these eigenstates, the fermionic annihilation operator is 
represented as 
%%%%%%%%%%%%%%%%%%%%%%%%%%%%%%%%%%%%%%%%%%%%%%%%%%%
\begin{eqnarray}
c_{m\sigma}&=&\sum_{\alpha,\alpha'}
D^{m\sigma}_{\alpha \alpha'}|\alpha \rangle \langle \alpha'|,\\
D^{m\sigma}_{\alpha \alpha'}&=&\langle \alpha|c_{m\sigma}|\alpha' \rangle.
\end{eqnarray}
%%%%%%%%%%%%%%%%%%%%%%%%%%%%%%%%%%%%%%%%
 
We now introduce the ionic resolvent and its self-energy, which are
represented in terms of the basis set $|\alpha \rangle$ as,
%%%%%%%%%%%%%%%%%%%%%%%%%%%%%%%%%%%%%%%%%%%
\begin{eqnarray}
&R_{\alpha}(\omega)&=[\omega -E_{\alpha}-\Sigma_{\alpha}(\omega)]^{-1}, \\
&\Sigma_{\alpha}(\omega)&=\sum_{m\sigma}\sum_{\alpha'}
(|D^{m\sigma}_{\alpha \alpha'}|^{2}+|D^{m\sigma}_{\alpha' \alpha}|^{2}) \nonumber \\
&\times \int^{\infty}_{\infty}&{\rm d}\epsilon 
\,f(\eta_{\alpha \alpha'}\epsilon)
\Delta_{m\sigma}(\epsilon)R_{\alpha'}(\omega +\eta_{\alpha \alpha'}\epsilon),
\label{eqn:SENCA}
\end{eqnarray}
%%%%%%%%%%%%%%%%%%%%%%%%%%%%%%%%%%%%%%%%%%%%%%%%%%%%%%%%
with  the Fermi distribution function
$f(\epsilon)$, where $\eta_{\alpha \alpha'}$ takes
 $1(-1)$ if the particle number of the state
$|\alpha \rangle $ is larger (smaller) than that of $|\alpha' \rangle $.
In eq.(\ref{eqn:SENCA}), we have neglected the vertex corrections, 
as often done in the treatment of DMFT.
The {\it full} Green function is now given by 
%%%%%%%%%%%%%%%%%%%%%%%%%%%%%%%%%%%%%%%%%5
\begin{eqnarray}
G_{m\sigma}(\omega)=\sum_{\alpha \alpha'}
|D^{m\sigma}_{\alpha \alpha'}|^{2}{\tilde G}_{\alpha \alpha'}(\omega),
\end{eqnarray}
%%%%%%%%%%%%%%%%%%%%%%%%%%%%%%
where ${\tilde G}_{\alpha \alpha'}(\omega)$ 
is written in terms of the ionic resolvents, 
%%%%%%%%%%%%%%%%%%%%%%%%%%%%%
\begin{eqnarray}
&&{\tilde G}_{\alpha \alpha'}(\omega)= 
\frac{1}{Z}\int {\rm d}\epsilon  \,{\rm e}^{-\beta \epsilon} \nonumber \\
&\times&[\rho_{\alpha}(\epsilon)R_{\alpha'}(\omega +\epsilon)
-\rho_{\alpha'}(\epsilon)R_{\alpha}(\epsilon -\omega)]
\label{eqn:NCAG}
\end{eqnarray}
%%%%%%%%%%%%%%%%%%%%%%
and
%%%%%%%%%%%%%%%%%%%%%%
\begin{eqnarray}
Z=\sum_{\alpha}\int {\rm d}\epsilon \,
{\rm e}^{-\beta \epsilon}\rho_{\alpha}(\epsilon).
\label{eqn:PF}
\end{eqnarray}
%%%%%%%%%%%%%%%%%%%%%%%%%%%%%%%%%%%%%%%%%%%%%%%%%%%%%
In eq. (\ref{eqn:NCAG}) and (\ref{eqn:PF}), 
$\rho_{\alpha}(\omega)=-1/\pi {\rm Im}(R_{\alpha}(\omega))$ 
is the spectral function of the ionic resolvent.

This completes the self-consistent procedure for DMFT with NCA.

%%%%%%%%%%%%%%%%%%%%%%%%%%%%%%%%%%%%%%%%
\section{Numerical Results}
%%%%%%%%%%%%%%%%%%%%%%%%%%%%%%%%%%%%%%%%
We numerically iterate the procedure in the 
previous section until the calculated quantities  converge  
within desired accuracy. In the following discussions, 
the bandwidth $D$ is taken to be unity for simplicity.

%=======================================
\subsection{Heavy quasi-particles near quarter filling}
%=======================================

\begin{figure}[h]
\vspace{-5mm}
\epsfxsize=9cm
\centerline{\epsfbox{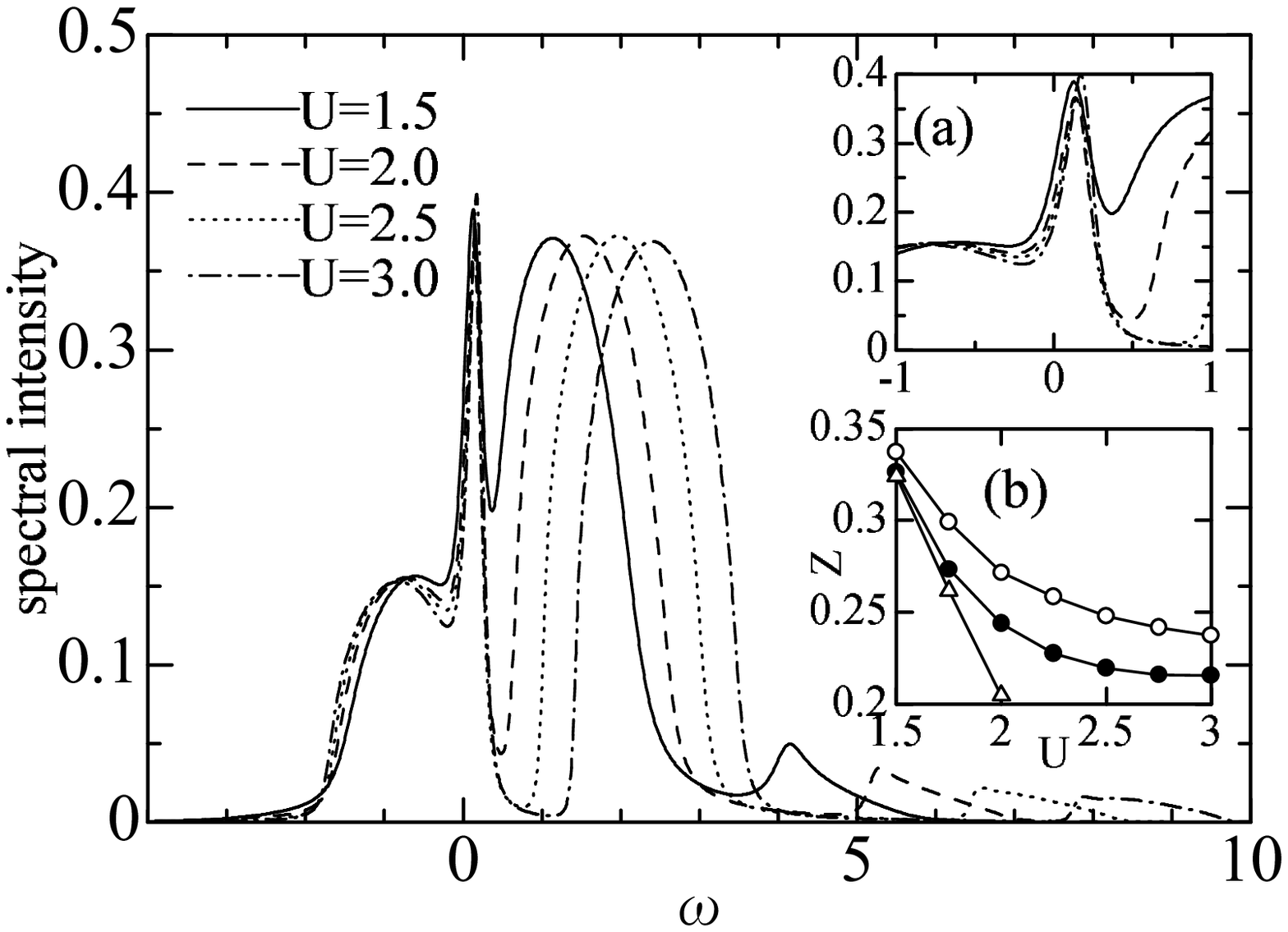}}
\vspace{-60mm}
\caption{One-particle spectral function $\rho_{m\sigma}(\omega)$ 
at $T=0.1$ for $J=0$ and $\Delta=0$.
The electron density per site is $n_{\rm tot}=0.96$
(i.e. close to quarter filling).
 The energy is measured from the Fermi level $\omega=0$, 
and the bandwidth $D$ is taken to be unity.
The inset (a) shows the magnification of the low-frequency part,
while (b) shows the quasi-particle renormalization factor $Z$
as a function of $U$ for $n_{\rm tot}=0.96$ (open circles), 
$n_{\rm tot}=0.98$ (filled circles) and $n_{\rm tot}=1.00$ (triangles).
}
\label{fig:DOS96}
\end{figure}

Let us start with a metallic system 
close to quarter filling ($n_{\rm tot}\sim 1$).
 We calculate the one-particle spectral function defined by
%%%%%%%%%%%%%%%%%%%%%%%%%%%%
\begin{eqnarray}
\rho_{m\sigma}(\omega)=-\frac{1}{\pi} {\rm Im}\,G_{m\sigma}(\omega)
\end{eqnarray}
%%%%%%%%%%%%%%%%%%%%%%%%%%%%%
for several different values of the Hubbard interaction $U$.
We first discuss the case in the absence of the Hund coupling, $J=0$.
The results are shown in Fig.\ref{fig:DOS96}.
%%%%%%%%%%%%%%%
% Fig. 1
%--------- DOS ($B&$(B=J=0) --------------------------------------------
%%%%%%%%%%%%%%%
As the Hubbard interaction increases, 
a sharp peak structure appears and gets narrower
around the Fermi level, which 
is caused by the formation of heavy quasi-particles.\cite{Kot96,Lom00,Ohk89}
At the same time, it is seen that 
a pseudo-gap structure develops just above the Fermi level
because the Hubbard interaction suppresses 
the charge fluctuation while enhances the spin fluctuation 
near quarter filling.
This behavior is similar to that expected for the hole-doped 
Hubbard model with single orbital.
The other peaks are easily identified with  the singly, 
doubly and triply occupied states, respectively.
The intensity of the fully occupied state is too small 
to observe in this figure.

The formation of a heavy quasi-particle band is 
more clearly seen in the quasi-particle renormalization factor 
shown in the inset (b) of Fig.\ref{fig:DOS96}, which is defined by 
%%%%%%%%%%%%%%%%%%%%%%%%%%%%
\begin{eqnarray}
Z={\biggl[}1-\frac{\partial \Sigma(\omega)}{\partial \omega}{\bigg|}_{\omega=0}
{\biggr]}^{-1}.
\end{eqnarray}
%%%%%%%%%%%%%%%%%%%%%%%%%%%
Since the mass enhancement factor of quasi-particles is 
inversely proportional to the renormalization factor, 
it is confirmed  that the mass of the renormalized 
band indeed increases as the Hubbard interaction $U$ increases.
%%%%%%%%%%%%%%%
% Fig. 2
%--------- DOS (U=2 n=0.96 $B&$(B=0) -------------------------------------
\begin{figure}[h]
\vspace{-5mm}
\epsfxsize=9cm
\centerline{\epsfbox{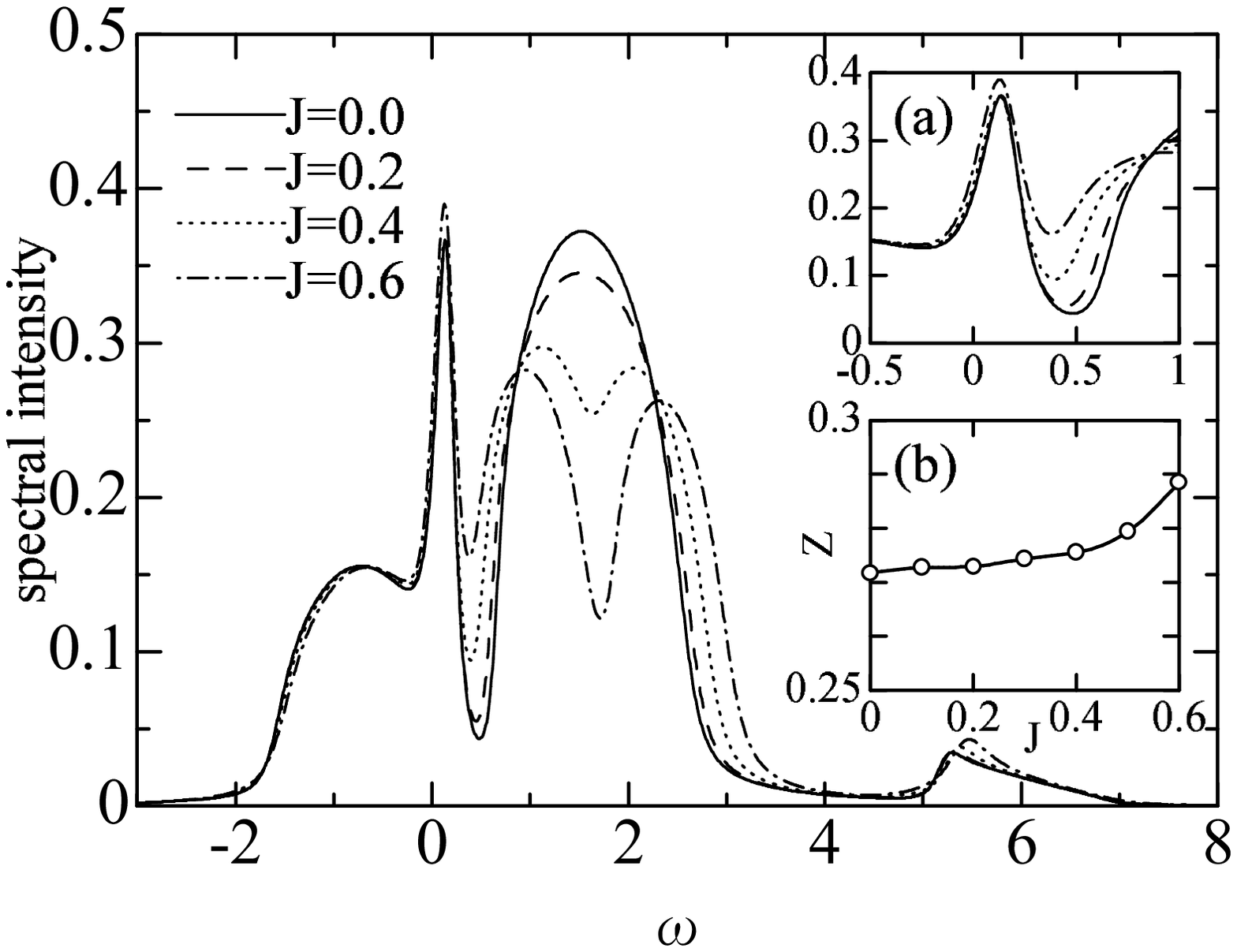}}
\vspace{-60mm}
\caption{One particle spectral function for various $J$ at $T=0.1$ and $U=2$.
The electron concentration is $0.96$.
The insets (a) and (b) show the magnification of the low-frequency behavior
and the renormalization factor.
}
\label{fig:DOSJ96}
\end{figure}
%%%%%%%%%%%%%%%
In Fig. \ref{fig:DOSJ96}, we show how the Hund coupling $J$ affects the 
spectral properties near quarter filling.  Although
 the spectral weight for the doubly occupied 
state splits into two parts as $J$ increases, 
the quasi-particle spectrum near the Fermi level 
is not so much affected by weak or intermediate $J$.
In particular, for the strong Hubbard interaction,  
the Hund coupling hardly affects the quasi-particle mass
although it modifies the spectrum in the high energy regime.

\begin{figure}[h]
\vspace{-5mm}
\epsfxsize=9cm
\centerline{\epsfbox{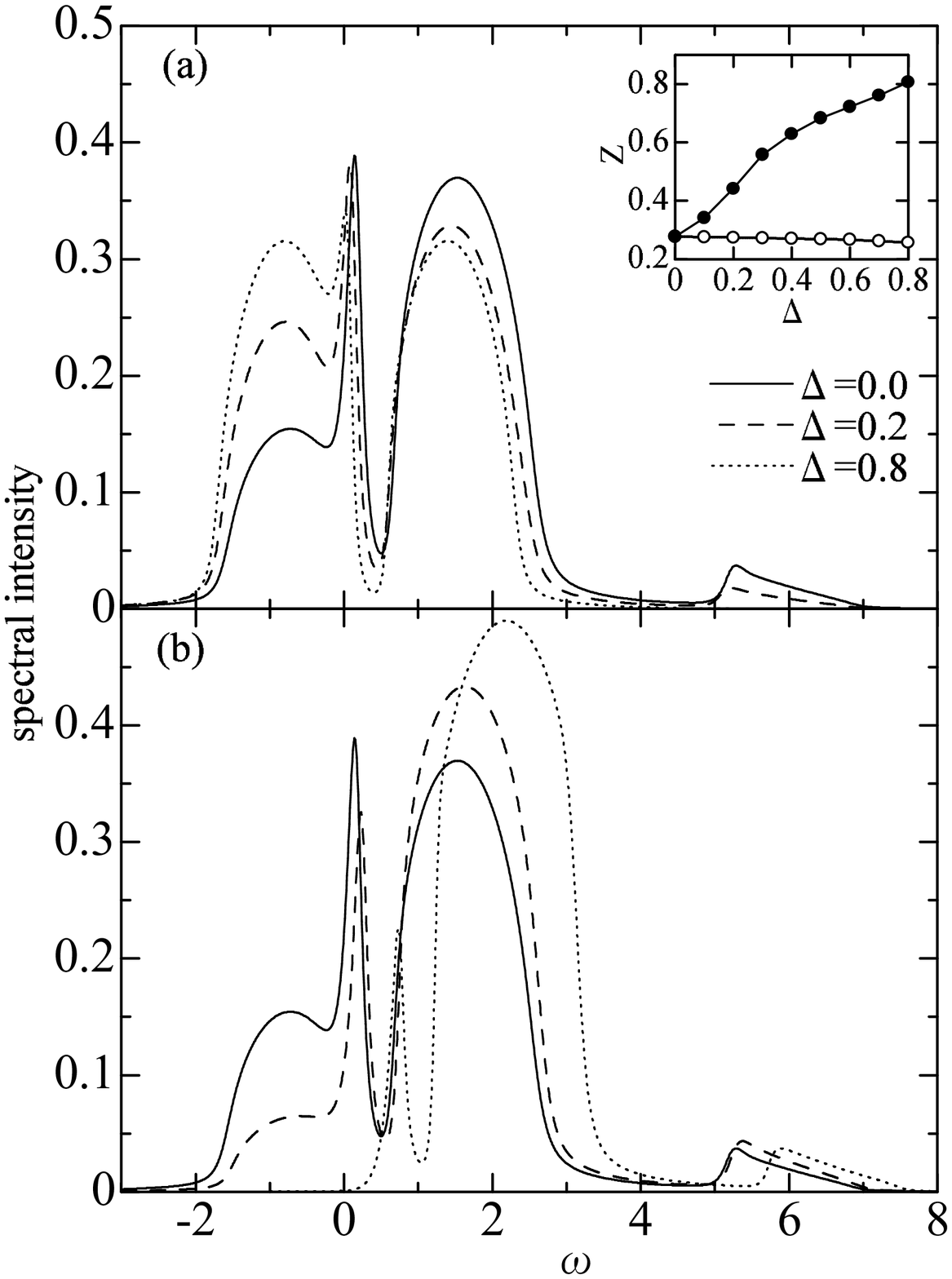}}
\vspace{-15mm}
\caption{One particle spectral function for 
 $U=2$, $J=0$ and $T=0.1$:
(a) lower band and (b) upper band.
The electron concentration is $n_{\rm tot}=0.96$.
The inset shows the renormalization factor 
for lower band (open circles) and upper band (filled circles).
}
\label{fig:DOSD96}
\end{figure}

Similarly, even when the splitting between two orbitals is introduced,
the formation of quasi-particles is not so much affected,
as seen from  Fig. \ref{fig:DOSD96}.
%%%%%%%%%%%%%%%
% Fig. 3
%--------- DOS (U=2 $B&$(B.ne.0) -----------------------------------------
%%%%%%%%%%%%%%%
For the lower band, the spectral weight is gradually transferred from 
 the doubly and triply occupied states to
 the singly occupied state as  $\Delta$ increases, 
while for the upper band, the weight for the singly
occupied state decreases, as should be expected.
Note that a quasi-particle band always exists in the 
lower band irrespective of $\Delta$. 
We can check that the renormalization factor for the 
lower band is almost unchanged as a function of $\Delta$.
(Fig. \ref{fig:DOSD96}, inset)
For sufficiently large $\Delta$, 
the spectrum of the lower band 
is reduced to that of the single-orbital
model near half filling, which is indeed
seen in the case of  $\Delta=0.8$.

Summarizing, the formation of heavy quasi-particles near quarter filling
is caused by the Hubbard interaction, as is the case 
for the single-orbital model. In particular, a heavy quasi-particle band and
a pseudo-gap structure are almost always formed for reasonably
 large $U$, irrespective of the strength 
of the orbital splitting and the Hund coupling.
As we will see momentarily,  this is not the case for 
the formation of heavy quasi-particles near half filling.

%=======================================
\subsection{Heavy quasi-particles near half filling}
%=======================================

\begin{figure}[h]
\vspace{-5mm}
\epsfxsize=9cm
\centerline{\epsfbox{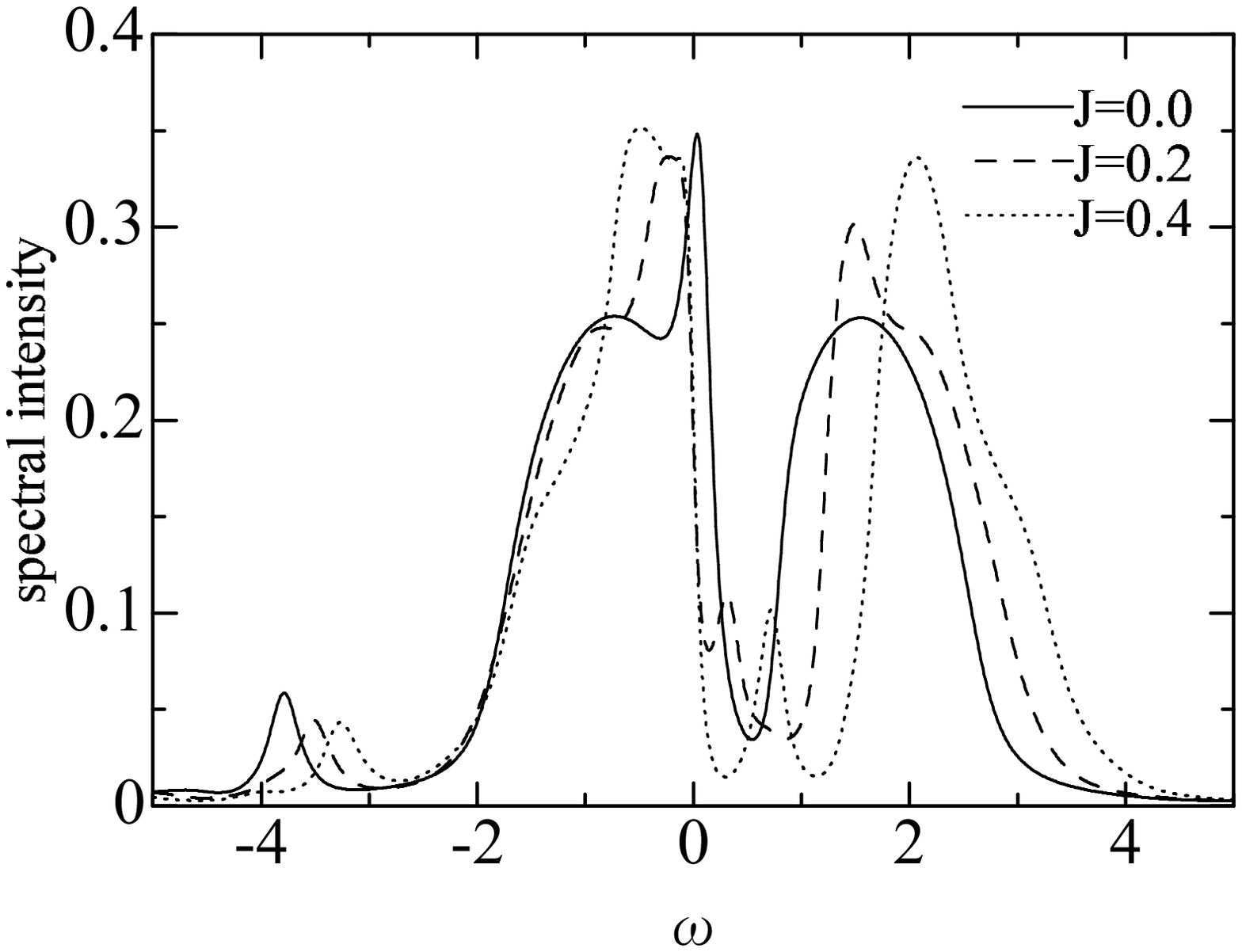}}
\vspace{-60mm}
\caption{One particle spectral function 
near half filling, $n_{\rm tot}=1.90$:
 $U=2$, $\Delta =0$ and $T=0.1$.
}
\label{fig:DOSJ190}
\end{figure}

We now turn to a metallic system  close to half-filling 
($n_{\rm tot}\sim 2$). 
Let us first discuss the system without orbital splitting.
Fig.\ref{fig:DOSJ190} shows the one-particle spectrum as a function of 
the Hund coupling $J$.
%%%%%%%%%%%%%%%
% Fig. 4
%--------- DOS (U=2 $B&$(B.ne.0 n=1.90) -----------------------------------------
%%%%%%%%%%%%%%%
As mentioned above, a heavy quasi-particle
 band exists around the Fermi level for $U=2$.
In contrast to the quarter filling case, 
the spectrum near the Fermi level is considerably affected by
the Hund coupling, 
since the energy eigenvalues of two-particle configuration 
are quite sensitive to the strength of the Hund coupling.
With the increase of $J$, the heavy quasi-particle band 
disappears and the charge excitation gap 
develops  around the Fermi level.
This implies that the Hund coupling has a tendency to
drive the system to the Mott insulator together with the 
Hubbard interaction \cite{Han98}, in contrast to the quarter-filling case.

\begin{figure}[h]
\vspace{-5mm}
\epsfxsize=9cm
\centerline{\epsfbox{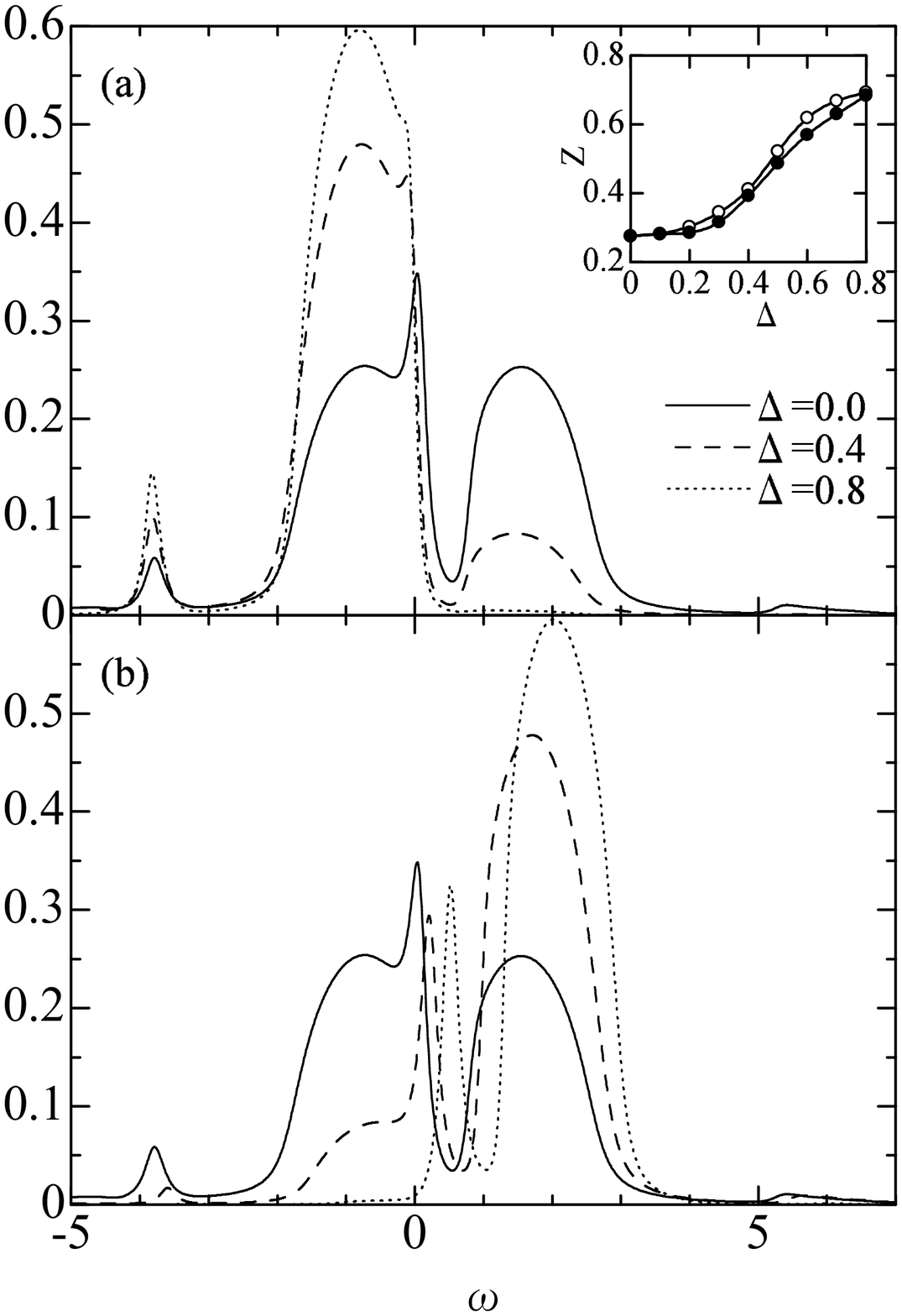}}
\vspace{-15mm}
\caption{One-particle spectral function for various values of $\Delta$:
$U=2$, $J=0$ and $T=0.1$.  The system is close to 
half filling, $n_{\rm tot}=1.90$.
(a) and (b) represent the spectrum for the lower band and 
the upper band, respectively.
The inset shows the renormalization factor 
for lower band (open circles) and upper band (filled circles).
}
\label{fig:DOSD190}
\end{figure}

We next discuss how the orbital splitting 
$\Delta$ affects the formation of heavy quasi-particles near half filling. 
We present the results for the spectral function 
near half-filling in Fig. \ref{fig:DOSD190}.
%%%%%%%%%%%%%%%
% Fig. 5
%--------- DOS (U=2 $B&$(B.ne.0 n=1.90) ----------------------------------
%%%%%%%%%%%%%%%
The overall structure in the spectrum has the shape expected naively.
Namely, with the increase of $\Delta$,
the weight of the middle peak (doubly occupied state) gets larger 
and that of the top peak (triply occupied state) becomes smaller in the
lower band case (a),
while in the upper band case (b), the opposite behavior is observed.
It should be noted  here that 
the quasi-particle band disappears gradually with the 
increase of $\Delta$ 
and is absorbed into the doubly occupied state.
This tendency is different from that observed for quarter 
filling. Near half-filling, both of the spin fluctuation and 
the charge fluctuation  are suppressed  in the presence of $\Delta$, 
since the lower band is almost fully occupied by electrons 
of spin singlet.
As a result, the system behaves like a renormalized band insulator with 
both of the spin and charge gaps. Therefore,  
the formation of heavy electrons is suppressed by  $\Delta$
in a metallic phase close to half filling.

\begin{figure}[h]
\vspace{-5mm}
\epsfxsize=9cm
\centerline{\epsfbox{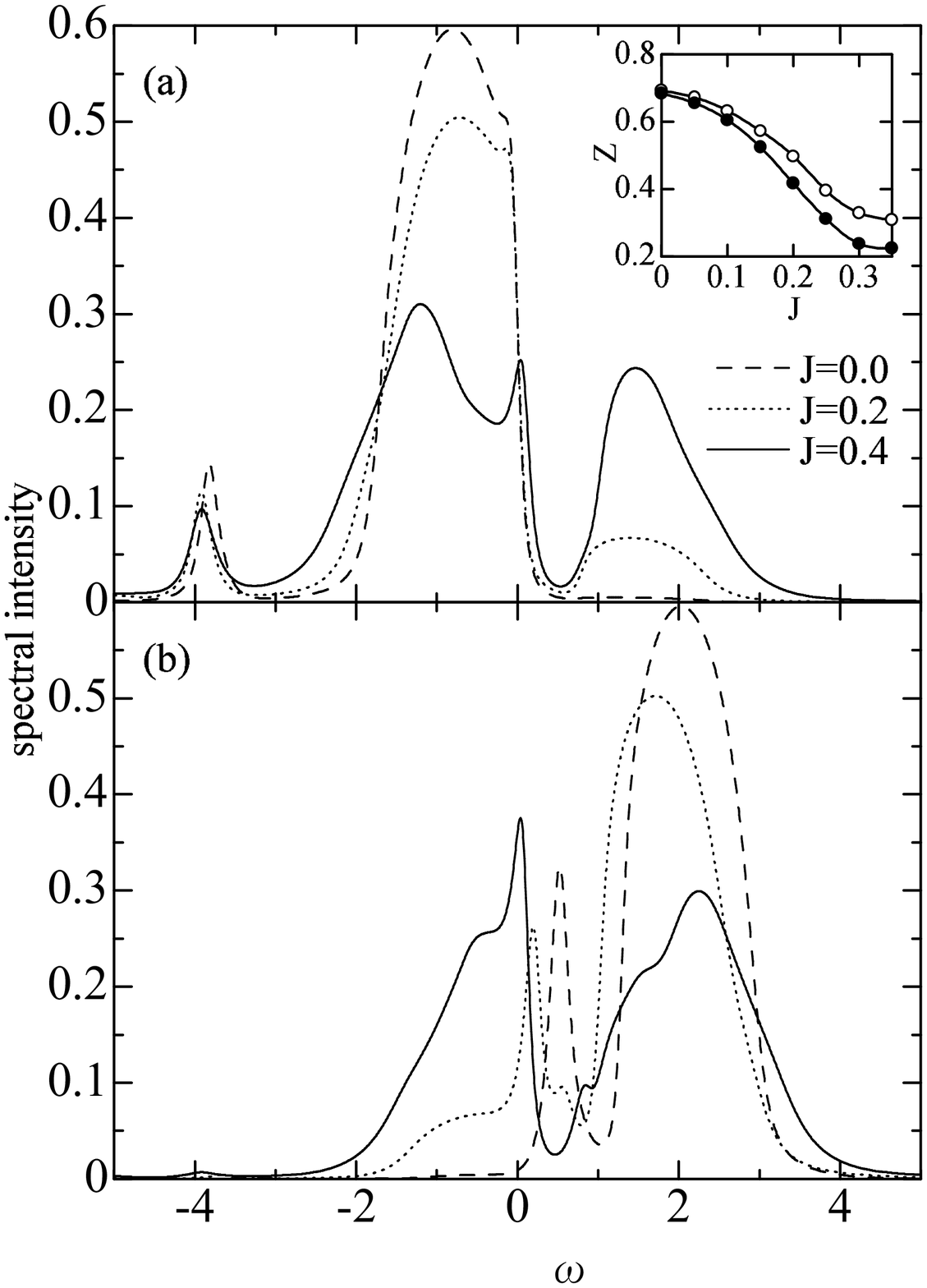}}
\vspace{-15mm}
\caption{One-particle spectral function 
at $U=2$, $\Delta=0.8$ and $T=0.1$.
The electron concentration is 1.90 (near half filling).
(a) and (b) represent the spectrum for the lower band and 
the upper band, respectively.
The inset shows the renormalization factor 
for lower band (open circles) and upper band (filled circles).
}
\label{fig:DOSDJ190}
\end{figure}

The Hund coupling, however, gives rise to a dramatic change in the 
spectrum in this case.
Fig. \ref{fig:DOSDJ190} shows the spectral function 
for $\Delta=0.8$ in the presence of the Hund coupling.
%%%%%%%%%%%%%%%
% Fig. 6
%--------- DOS (U=2 $B&$(B,J.ne.0) --------------------------------------------
%%%%%%%%%%%%%%%
Although a quasi-particle band around the Fermi level 
is almost invisible in the absence of the Hund coupling ($J=0$), 
it is enhanced again with the increase of $J$ in both (a) and (b).
In the absence of the Hund coupling, each lattice site
is mostly occupied by two electrons of spin singlet in the same orbital.
When the Hund coupling is introduced, the energy of  
the spin-triplet state
is lowered, and can be degenerate with the singlet
state. This induces the large spin fluctuation, enhancing  
the quasi-particle mass to form a heavy quasi-particle band.
This tendency is clearly seen in
the renormalization factor shown 
in Figs. \ref{fig:DOSD190} and \ref{fig:DOSDJ190}.
As seen in the inset of Fig. \ref{fig:DOSD190}, when the orbital splitting 
increases, the renormalization factor for both of the lower (open triangle) 
and  upper bands (filled triangle)  increases, while 
in the inset of Fig. \ref{fig:DOSDJ190} the introduction of the Hund coupling 
again reduces the renormalization factor. 
These results imply that the spin fluctuation 
caused by the Hund coupling enhances the quasi-particle mass
in a metallic state near half filling, which is not observed 
near quarter filling.

%=======================================
\subsection{Optical Conductivity}
%=======================================

\begin{figure}[h]
\vspace{-5mm}
\epsfxsize=9cm
\centerline{\epsfbox{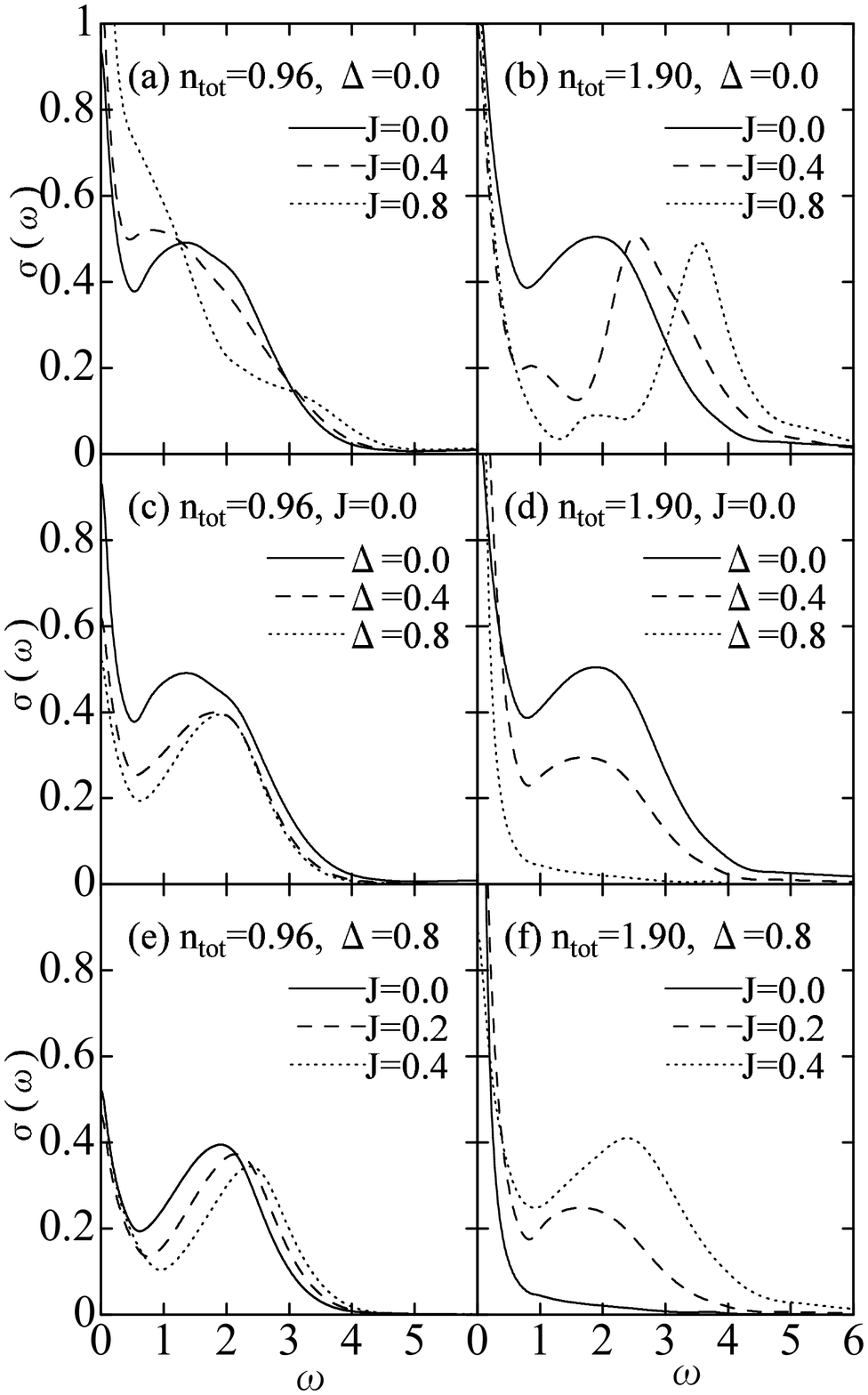}}
\vspace{-0mm}
\caption{Optical conductivity $\sigma (\omega)$ 
for (a) $\Delta =0$ and $n_{\rm tot}=0.96$, 
(b) $\Delta =0$ and $n_{\rm tot}=1.90$,
(c) $J=0$ and $n_{\rm tot}=0.96$,
(d) $J=0$ and $n_{\rm tot}=1.90$, 
(e) $\Delta =0.8$ and $n_{\rm tot}=0.96$, and
(f) $\Delta =0.8$ and $n_{\rm tot}=1.90$.
The other parameters are  $U=2$ and $T=0.10$.
}
\label{fig:OPT}
\end{figure}

In order to further investigate the dynamical response, 
we calculate the optical conductivity.
We employ the following formula for the optical conductivity, 
%%%%%%%%%%%%%%%%%%%%%%%%%%%%%%%%%%%%%%%%%%%%%
\begin{eqnarray}
&&\sigma(\omega)=\pi \sum_{m,\sigma} 
\int_{-\infty}^{\infty} {\rm d}\epsilon' 
\int_{-\infty}^{\infty} {\rm d}\epsilon \,N_{0}(\epsilon)\nonumber \\
&\times&A_{m\sigma}(\epsilon,\epsilon')
A_{m\sigma}(\epsilon,\omega+\epsilon')
\frac{f(\epsilon')-f(\epsilon'+\omega)}{\omega},\\
&A_{m\sigma}&(\epsilon,\omega)=-\frac{1}{\pi}
{\rm Im}{\Big (}\frac{1}{
\omega+{\rm i}\delta-\epsilon+\mu-\Sigma_{m\sigma}(\omega)}{\Big )},
\label{eqn:opt-con}
\end{eqnarray}
%%%%%%%%%%%%%%%%%%%%%%%%%%%%%%%%%%%%%%%%%%%%%%%%
where the vertex correction is neglected,
which is justified in the limit of large dimensions
\cite{Mul89,Khu90,Pru93}.
The computed results are shown in Fig. \ref{fig:OPT}.
%%%%%%%%%%%%%%%
% Fig. 7
%--------- Optical Conductivity --------------------------------------
%%%%%%%%%%%%%%%
In contrast to the one-particle spectrum, the 
overall structure of the optical conductivity is not so sensitive
to the parameters employed.  Nevertheless, we can see some 
distinct behaviors near quarter filling and half filling.
We first discuss the orbitally degenerate case 
($\Delta=0$) near quarter filling and half filling 
shown in Fig. \ref{fig:OPT} (a) and (b). 
It is clearly seen both in (a) and (b)  for $J=0$ that
 the optical conductivity consists of the Drude-like part 
($\omega \sim 0$) and the interband part.
With the increase of $J$ in (a), the weight for
the interband absorption gradually shifts to the lower-energy regime, 
reflecting  the splitting of 
the  doubly occupied state due to the Hund coupling,
as already shown in Fig. \ref{fig:DOSJ96} (a).
On the other hand in (b) the Drude-like part gets narrower and 
the interband absorption shifts to the higher-energy part 
with increase of $J$.  This result in (b) reflects that 
the Hund coupling has a tendency to enhance
 the charge excitation gap \cite{Han98},
which is different from the case near quarter filling.

We next discuss the effect of the orbital splitting by observing 
Fig. \ref{fig:OPT} (c) and (d), where we assume that 
the  Hund coupling is absent for simplicity. As seen from (c), 
near quarter filling, 
the amplitude of the conductivity becomes smaller 
with the increase of $\Delta$  although  the overall structure 
is almost unchanged. For sufficiently large $\Delta$, 
the amplitude becomes about half of the case of $\Delta=0$
since the lower band corresponds to that of the single-band model and 
the upper one exists above the Fermi level \cite{Kot96}
(see Fig. \ref{fig:DOSD96}).
On the other hand,  the Drude-like peak in (d) becomes 
narrower while the interband-absorption 
weight gets smaller as $\Delta$ increases because
both of the charge and spin fluctuations are suppressed 
for large $\Delta$ near half filling, as seen 
in Fig. \ref{fig:DOSD190}. Therefore, 
a slight hole-doping into half-filled band brings about the sharp 
Drude-like absorption peak in the low energy part.

Finally, we discuss the interplay of the Hund coupling and the 
orbital splitting. As seen in Fig. \ref{fig:OPT} (e),
 the conductivity does not change its characteristic
behavior except that 
the interband absorption slightly shifts to the higher energy 
regime with the increase of $J$.  This is because 
the Hund coupling  hardly affects the low-energy
spectrum for large $\Delta$ near quarter filling,
since the lower band is reduced to that of the single-band 
Hubbard model as mentioned in Fig. \ref{fig:DOSD190}.
Near half filling (Fig. \ref{fig:OPT}(f)), however,
a rather drastic change occurs in the conductivity:
the Drude-like part gets wider 
and the interband absorption is induced again as $J$ increases,
reflecting the appearance of a 
quasi-particle state near the Fermi level 
as shown in Fig. \ref{fig:DOSDJ190}.

%%%%%%%%%%%%%%%%%%%%%%%%%%%%%%%%%%%%%%%%
\section{Summary and Discussions}
%%%%%%%%%%%%%%%%%%%%%%%%%%%%%%%%%%%%%%%
We have studied how the interplay of the Hubbard interaction, 
the Hund coupling and the orbital splitting affects the formation 
of  heavy quasi-particles in the two-orbital  Hubbard model.
In order to investigate dynamical quantities in the whole energy region, 
we have exploited dynamical mean field theory with
the non-crossing approximation.  We
 have discussed  the one-particle spectral function 
and the optical conductivity for a
metallic system in the vicinity of the Mott insulator.
In particular,  similarity and difference in the formation of 
heavy quasi-particles between quarter filling and half filling have been
clarified. Near quarter filling, hole-doping into the 
Mott insulator leads to the formation of the 
heavy quasi-particle band, which is 
slightly renormalized by the Hund coupling or the orbital splitting.
On the other hand, near half filling
the orbital splitting prevents the formation of 
heavy particles, driving to the system to a correlated
band-type insulator, for which  a heavy quasi-particle band is not 
formed by hole doping.  If 
the Hund coupling is switched on, however,
heavy quasi-particles appear again due to the 
enhanced spin fluctuation.

In connection with the experimental results for transition 
metal oxides such as 
${\rm La_{1-{\it x}}Sr_{\it x}TiO_{3}}$, \cite{Ima98}
the quarter-filled model ($n_{\rm tot}\sim 1$) 
has been studied extensively.
\cite{Ani97,Zol00,Nek00}
A heavy quasi-particle band studied here was pointed out in a
different method, and  the overall structure 
of the one-particle spectrum was quantitatively 
discussed in comparison  with the photoemission
experiments,\cite{Fuj92}
 although a sharp quasi-particle band in the 
spectrum  was not clearly observed experimentally.
On the other hand, the half-filled Hubbard model with two orbitals
has not been studied systematically thus far in comparison with 
the quarter-filled model.
A possible example in transition metal
oxides relevant for this case ($n_{\rm tot}\sim 2$) may be 
${\rm La_{1-{\it x}}Sr_{\it x}VO_{3}}$ \cite{Ina95,Ima98} 
with electron configuration of $3d^{2}$ in V ions.
In this material, the Hund coupling forms the $S=1$ spin at 
each site, resulting in the antiferromagnetic ground state 
at half filling. According to the results in \S 3.2 (Fig.4), when holes are 
doped into this system to realize a paramagnetic metal,
 a heavy quasi-particle band may not be observed\cite{Ina95}
because of a rather large Hund-coupling.
 Nevertheless, we think that for related 
materials where the orbital splitting due to a crystalline field 
becomes comparable to the Hund coupling, a heavy quasi-particle formation 
may be more easily observed experimentally, reflecting  the 
interplay of the above two effects, as  pointed out in Fig.6.

%%%%%%%%%%%%%%%%%%%%%%%%%%%%%%%%%%%%%%%%
\section*{Acknowledgements}
We would like to dedicate this paper to Prof. E. M$\ddot{\rm u}$ller-Hartmann 
on occasion of his 60th birthday.
We acknowledge valuable discussions with Th. Pruschke.
The work is partly supported by a Grant-in-Aid from the Ministry of 
Education, Science, Sports, and Culture. 
Parts of the numerical computations were done by the supercomputer center 
at the Institute of the Solid State Physics, The University of Tokyo.

\end{document}